\documentclass[epj,amsfonts]{svjour}
\usepackage{graphics}
\usepackage{graphicx}
\usepackage{dcolumn}
\usepackage{amssymb}
\usepackage{amsmath}
\usepackage{color}
\usepackage{widetext}
\newcommand{\ar}{\arrowvert}

\newcommand{\cd}{\! \cdot \!}
\newcommand{\be}{\begin{equation}}
\newcommand{\ee}{\end{equation}}
\newcommand{\ba}{\begin{eqnarray}}
\newcommand{\ea}{\end{eqnarray}}

\begin{document}
\title{Bulk viscosity and energy-momentum correlations\\
 in high energy hadron collisions}
\author{ 
Antonio Dobado, Felipe J. Llanes-Estrada and Juan M. Torres-Rincon
}                     
\institute{Departamento de F\'{\i}sica Te\'orica I,  Universidad
Complutense, 28040 Madrid, Spain}
\date{Received: date / Revised version: date}
%

\abstract{
We show how the measurement of appropriately constructed particle-energy/momentum correlations allows access to the bulk viscosity of strongly interacting hadron matter in heavy ion collisions.\\
This measurement can be performed by the LHC and RHIC experiments in events with high-particle multiplicity, following up on existing estimates of the shear viscosity based on elliptic flow. 
} 
\authorrunning{Dobado, Llanes-Estrada and Torres-Rinc\'on}
\titlerunning{Bulk viscosity and momentum correlations}
\maketitle

\section{Introduction}

A relatively broad consensus arose in the last years around the nature of the medium created in heavy ion collisions when it has become established (to the surprise of many) that the medium created at energy densities beyond the phase transition to a quark-gluon medium is best characterized as a strongly interacting fluid.  This fluid features a very low shear viscosity over entropy density $\eta/s\simeq 0.1$, that is, a liquid close to perfect where neighbooring fluid elements strongly couple (although some alternative possibilities such as plasma instabilities could still allow for a weakly coupled medium~\cite{Mrowczynski:2009gf}).  This apparent low viscosity has been confirmed recently at the LHC~\cite{Lacey:2010ej}. Once clear that the medium is behaving as a fluid, one would like to explore all other transport coefficients.\\

Bulk viscosity is of current interest for two reasons. The first is that, near the phase transition, the condition $\eta\gg \zeta$ of usual fluid mechanics, that allows one to neglect the bulk viscosity is probably not valid in heavy ion collisions. Indeed it has been pointed out that
$\eta/s$ may have a minimum at the phase transition~\cite{Csernai:2006zz} although
it is not clear that the minimum must lay at precisely this transition in an arbitrary system, as shown by the counterexample of the linear sigma model in large $N$~\cite{Dobado:2009ek}. Moreover, near the phase transition the bulk viscosity might have a maximum~\cite{Karsch:2007jc,FernandezFraile:2008vu}, although again in certain models such as the one-dimensional Gross-Neveu model the bulk viscosity is monotonously decreasing~\cite{FernandezFraile:2010gu}. \\
Notwithstanding the uncertainties, the scenario of a small shear and a relatively large bulk viscosity remains possible near the quark-gluon liquid phase transition, making the volume viscosity more important than previously thought from phase-shift based hadron evaluations~\cite{Davesne:1995ms}.

The second reason is the obvious realization that the bulk viscosity controls the relaxation of the trace of the stress-energy tensor $T^\mu_\mu$ when it separates from equilibrium, and is therefore related to dilatation invariance.
Since this is a symmetry of the classical QCD Lagrangian (in the chiral limit),
bulk viscosity will become a probe of the quantum trace anomaly that erases the symmetry from physical amplitudes and spectral properties (such as the proton mass).

The most popular way of estimating the shear viscosity, by means of elliptic flow, is by its very definition however not useful for the measurement of the bulk viscosity $\zeta$ (we will make no distinction between bulk, volume and second viscosity). \\
Alternative ways of accessing the second viscosity have been proposed~\cite{Torrieri:2007fb} but, in our view, the most promising avenue for a timely estimate is provided by the method of particle momentum correlations/fluctuations, that we study in this article, and that has been put to use to test models of the collision~\cite{Cunqueiro:2007zz}. The analogous construction for the measurement of shear viscosity has been developed by Gavin and Abdel-Aziz
~\cite{Gavin:2007zz,Gavin:2006xd} and there are already preliminary measurements from STAR that indicate broad agreement between this method and the elliptic flow estimates~\cite{Sharma:2009zt}.

\section{Fluctuations of the Stress-Energy tensor}
 An active degree of freedom needs to maintain an energy of order $k_B T$ at thermal equilibrium characterized by temperature $T$, but this energy dissipates according to transport equations in the medium. Hence spontaneous statistical fluctuations have to occur in the system at the right rate to reexcite the dissipated energy to the thermal level.
This is the intuitive explanation of the fluctuation-dissipation theorem~\cite{Callen:1951vq}, as a necessity of energy equipartition.
Thus, one expects the bulk viscosity to be proportional to the fluctuations of the stress-energy tensor, and particularly of its trace. 

If one divides the stress-energy tensor into an ideal and an out-of-equilibrium part $\Theta_{\mu\nu}=T_{\mu\nu}+\tau_{\mu\nu}$, the latter can be expressed for a fluid element at rest but with velocity gradients not necessarily vanishing, as~\cite{Landaufluids}
\ba \label{separatensores}
\tau_{ik} = -\eta\left(v_{i,k}+v_{k,i}-\frac{2}{3}\delta_{ik}v_{l,l} \right)
- \zeta v_{l,l} \delta_{ik} + t_{ik} \\ \nonumber
\tau_{00}=\tau_{0i}=0 .
\ea
 This formula features both bulk $\zeta$ and shear viscosity $\eta$, and in addition to the hydrodynamical contribution proportional to velocity gradients, a fluctuating part of purely statistical origin $t_{ik}$.

To experimentally access the bulk viscosity, we want to exploit the connection between the two-point correlation function of this fluctuating contribution to the stress-energy tensor $t_{ik}$, and the viscosities, demonstrated also by Landau and Lifschitz~\cite{Landaufluids}, as
\ba
\langle t_{ik}({\bf r}_1,t_1) t_{lm}({\bf r}_2,t_2)\rangle = 
\delta({\bf r}_1-{\bf r}_2)\delta(t_1-t_2) \times \\ \nonumber
2T \left[ \eta \left( \delta_{il} \delta_{km} + \delta_{im} \delta_{kl} \right) + \left(\zeta - \frac{2}{3} \eta\right) \delta_{ik} \delta_{lm}\right]  
\ea
(here $T$ is the temperature).

To separate the bulk viscosity one needs to take the trace over the spatial indices so that
\be \label{stresstovisco}
\langle t_{ii}({\bf r}_1,t_1) t_{ll}({\bf r}_2,t_2)\rangle = 
 18 T\ \zeta\ 
\delta({\bf r}_1-{\bf r}_2)\delta(t_1-t_2) \ .
\ee
Thus, if an experiment in heavy-ion collisions could access the correlations of the traced stress-energy tensor fluctuations, this would allow access to the desired bulk viscosity in a transparent manner. 

The route taken by Gavin and Abdel-Aziz~\cite{Gavin:2006xd} to access the shear viscosity is not too different, but somewhat simpler since only one element of the stress-energy tensor needs to be measured, for example $\tau_{0r}$ in cylindrical coordinates. This allows the extraction to be possible with only $p_t$ correlations. Instead, the bulk viscosity requires reconstruction of the trace of the tensor, and even in a perfectly central heavy ion collision where there is no vorticity and $T_{\phi\phi}$ can be neglected, one still needs to measure at least three correlations as will be shown shortly. This makes the measurement more challenging, but closed formulae can be found without more conceptual difficulty.

Let us first consider the case of a fluid element near rest so we can ignore relativistic corrections, and adopt cylindrical coordinates ${\bf x}'\equiv (r,\phi,z)$ with the $z$ direction along the collider beam axis. The Cartesian tensor transforms under this coordinate change ${\bf x}\to {\bf x}'$ as
\be
\tau^{'ij} = \frac{\partial x^{'i}}{\partial x^k} 
\frac{\partial x^{'j}}{\partial x^l}  \tau^{kl} \ .
\ee
The trace of the tensor is then
\be
\tau^{xx}+\tau^{yy}+\tau^{zz} = \tau^{rr} + r^2 \tau^{\phi\phi} + \tau^{zz}
\ee
in cylindrical coordinates. Now, if we assume that there is cylindrical symmetry for central collisions, one expects that $\partial {\bf v}/\partial \phi=0$ for the velocity (or any quantity whose azimuthal derivative is taken), meaning that $\tau^{\phi\phi}=0$. 

Thus we are left with
\ba \label{tracecorr}
\langle \tau_{ii}({\bf r}_1,t_1) \tau_{ll}({\bf r}_2,t_2)\rangle = \\ \nonumber
\langle (\tau_{rr}+\tau_{zz})({\bf r}_1,t_1) (\tau_{rr}+\tau_{zz})({\bf r}_2,t_2)\rangle 
\ea
and as advanced this requires the measurement of three correlation components, due to the crossed terms, which will lead to the necessity of full energy reconstruction.

\section{Particle momentum fluctuations}

In this section we will show how the correlation of the fluctuation of the stress-energy tensor can be related to directly measurable particle momenta. 
Our discussion here closely parallels that of Gavin and Abdel-Aziz~\cite{Gavin:2007zz,Gavin:2006xd}, with the modifications appropriate for the bulk instead of the shear viscosity.

\subsection{Average Equilibrium Hypothesis}\label{AvgEq}

In heavy-ion collisions the experiment (e.g. a head-on lead-lead collision at the LHC) is repeated numerous times, so that a large database of equally prepared systems is formed. The $\langle\ \rangle$ average symbol is then understood as an average over all the recorded central collisions (those with large particle multiplicity are in common practice assumed to be head-on and therefore, more central).

The fundamental hypothesis underlying the analysis is that a state of hydrodynamic equilibrium is reached after the collision. This is supported by a large body of data from the RHIC experiments and is widely assumed to be a good approximation to reality. \\
The analysis we present requires a slight extension of this assumption. 
We require, as in~\cite{Gavin:2007zz}, that the average taken over all collisions coincides with the equilibrium state, that is, the dissipative part of the stress-energy tensor averages to zero
$\langle \tau_{ij} \rangle = 0$. This excludes systematic deviations from equilibrium that may affect all collisions, and in effect attributes deviations from equilibrium to event by event fluctuations. We refer to this as the ``Average Equilibrium Hypothesis''.
With this caveat in mind, we will obtain next in subsection~\ref{subsec:correl} the relation between the fluctuations of the stress-energy tensor and those of particle momenta.

As a last comment on this hypothesis, let us note that it is in the spirit of the Gibbs ensemble and the ergodic hypothesis, where the study of many copies of the same system at fixed time is equivalent to the study over
very large times of the fluctuations of one given system. 

This large time also appears in Kubo's formula for the bulk viscosity
\be
\zeta = \frac{1}{9} \lim_{\omega\to 0} \frac{1}{\omega} \int_{-\infty}^\infty dt
\int d{\bf x} e^{i\omega t} \langle \left[
T_{ii}(t,{\bf x}), T_{jj}(0)
\right]\rangle
\ee
that makes natural to employ Eq.~(\ref{stresstovisco}) as a starting point to access the viscosity.

\subsection{Correlation of stress-energy tensor $\langle \tau \tau \rangle$ and particle momenta.} \label{subsec:correl}

Let us concentrate on the correlation $\langle \tau_{rr} \tau_{rr}\rangle$
in Eq.~(\ref{tracecorr}), the extension to include the $zz-zz$ correlation and the $rr-zz$ cross-correlation being trivial.

The number of particles per unit of phase space is, in terms of the distribution function,
\be \label{fdistribucion}
dn = f d^3 x \frac{d^3p}{(2\pi)^3} = (f_{\rm{eq}}+\delta f) d^3 x \frac{d^3p}{(2\pi)^3} 
\ee
where we split the off-equilibrium part of the distribution function $\delta f$.

It is easiest to anticipate the answer and consider the following particle-momentum correlation
\be
{\mathcal{C}}_{rr} = \langle \sum_{{\rm all\ }ij} \frac{(p_{ri})^2}{E_i} \frac{(p_{rj})^2}{E_j} \rangle \ .
\ee
In this formula, the sum over $i,j$ extends over all pairs of particles in a given collision event (this includes the square of the function for each particle). After this sum is performed for each individual event, the average is taken over all the events of the data sample.
Employing Eq.~(\ref{fdistribucion}) we obtain
\ba
\nonumber
{\mathcal{C}}_{rr} &=&  \int dn_1 dn_2  \langle   \frac{(p_{r1})^2}{E_1} \frac{(p_{r2})^2}{E_2} \rangle \\
&=& \int\frac{d^3p_1}{(2\pi)^3}\frac{d^3p_2}{(2\pi)^3} d^3x_1 d^3 x_2 \nonumber 
\langle   \frac{(p_{r1})^2}{E_1} \frac{(p_{r2})^2}{E_2} 
f(1,2) \rangle
 \\
&=& \int\frac{d^3p_1}{(2\pi)^3}\frac{d^3p_2}{(2\pi)^3} d^3x_1 d^3 x_2 \nonumber \\ 
& &
\langle   \frac{(p_{r1})^2}{E_1} \frac{(p_{r2})^2}{E_2} 
(f_{\rm{eq}1}+\delta f_1)(f_{\rm{eq}2}+\delta f_2) \rangle \ 
\ea
where the usual Boltzmann approximation of kinetic theory (valid for moderate densities) entails a factorization of the two-particle distribution function $f(1,2)=f_1\times f_2$.

Additionally invoking the Average Equilibrium Hypothesis of subsection~\ref{AvgEq} we can ignore the terms linear in 
$\delta f$ since $\langle \delta f \rangle =0$ (no systematic out-of-equilibrium effects). 
The term independent of $\delta f$ and the quadratic term contribute respectively the first and second lines of
\ba \label{Crr}
{\mathcal{C}}_{rr} &=& \langle N \rangle^2 \langle \frac{p_r^2}{E} \rangle^2  \nonumber \\
&+& \int d^3x_1 d^3 x_2 \langle \tau_{rr}({\bf x}_1) \tau_{rr}({\bf x}_2)
\rangle
\ea
with $N$ the total particle multiplicity in an individual event (preferably a large number, up to several thousand particles at the LHC, to select central collisions).

This equation relates the fluctuations of the stress-energy tensor to the particle momenta as measured in a detector, but it includes both the stochastic force $t_{ij}$ whose correlator reveals the viscosity in Eq.~(\ref{stresstovisco}), and the hydrodynamic part $\tau^{hyd}_{ij}$, the first terms in Eq.~(\ref{separatensores}). 
One can substitute directly $\tau_{ij}$ for $t_{ij}$ in the terms linear in $t$
\be
\langle \tau_{rr}({\bf x})\rangle = \langle t_{rr}({\bf x}) \rangle
= \langle \int \frac{d^3p}{(2\pi)^3}   \frac{(p_r)^2}{E} \delta f \rangle
= 0
\ee
under the Average Equilibrium 
Hypothesis~\footnote{As an aside, and since $\langle \tau_{rr} \rangle = 0$,
 one has $\int \langle \tau_{rr} \tau_{rr}\rangle = \int \left( \langle \tau_{rr} \tau_{rr}\rangle
- \langle \tau_{rr} \rangle \langle \tau_{rr} \rangle \right)$. Therefore one can equally talk of 
$\Delta \tau_{rr}$ if any conceptual advantage could be obtained.}.

\subsection{Separation of the stochastic stress-energy contribution $t_{ii}$}

This leaves the quadratic terms $\langle \tau^{hyd}_{rr}({\bf x}_1,t_1) \tau^{hyd}_{rr}({\bf x}_2,t_2)\rangle$ \\
and $\langle t_{rr}({\bf x}_1,t_1) t_{rr}({\bf x}_2,t_2)\rangle $. 
Gavin and Abdel-Aziz have been interested in the correlator of the hydrodynamic part, that satisfies a diffusion equation whose exponential solution has a characteristic diffusion time from which they can read off the shear viscosity, if one studies the long-time evolution of the system.
In this article we also propose to study the correlator of the stochastic force $t_{ij}$ whose intensity also gives access to the viscosity. The key to separate them is to perform a time-integration. 
Since Eq.~(\ref{stresstovisco}) contains a $\delta(t_1-t_2)$ (this just means that the stochastic force is statistically independent at each instant) whereas the hydrodynamic part features a soft time dependence
$e^{-\lambda t}$ in the time correlation function, an integration over very small time $\Delta {\mathcal{T}}$ picks up the stochastic part.

Indeed, for a general fluctuating magnitude satisfying
\be
\frac{dx}{dt} = - \lambda x + y
\ee
with a continuous time-correlation function 
\be
\langle x(t) x(t') \rangle = \langle x^2 \rangle e^{-\lambda \ar t-t' \ar}\ ,
\ee
the double integration
\be
\int_{\Delta {\mathcal{T}}}  dt_1  \int_{\Delta {\mathcal{T}}}  dt_2 \langle x(t) x(t') \rangle
= \langle x^2 \rangle (\Delta\mathcal{T})^2 + O\left(\Delta (\mathcal{T})^3\right).
\ee
is a second order infinitesimal. \\
However integrating over the stochastic part 
on the right hand side of Eq.~(\ref{stresstovisco}) 
(equivalent to the correlator $\langle y(t) y(t')$)
yields
\be
\int_{\Delta {\mathcal{T}}}  dt_1  \int_{\Delta {\mathcal{T}}}  dt_2
\delta(t_1-t_2) = \Delta\mathcal{T}
\ee
which is of first order.\\
Thus, to separate the statistically fluctuating and hydrodynamic parts we employ an integration over a short time interval. Under this integration one can exchange $\langle \tau \tau \rangle \to \langle t t\rangle$
in Eq.~(\ref{Crr}).

\subsection{Experimental observable in the fluid's rest-frame}

Finally we have the sought after relation between the correlations of the stress-energy tensor fluctuations and an observable in terms of particle momenta.
\ba \label{resexp}
\int_{\Delta {\mathcal{T}}}   \int_{\Delta {\mathcal{T}}} dt_1  dt_2
\int d^3x_1 d^3 x_2 \langle t_{rr}({\bf x}_1,t_1) t_{rr}({\bf x}_2,t_2)\rangle  \\ \nonumber
= (\Delta {\mathcal{T}})^2 \left(
\langle \sum_{{\rm all\ }ij} \frac{(p_{ri})^2}{E_i} \frac{(p_{rj})^2}{E_j} \rangle 
- \langle N \rangle^2 \langle \frac{p_r^2}{E} \rangle^2 \right)
\ea
where the average on the last term is a double average over both the event and the sample of events under certain kinematic cuts, multiplied by the average multiplicity, and the particles are supposed to have been emitted during the small interval $\Delta \mathcal{T}$ (how this can be seen experimentally is left for the next Sec.~\ref{secrelat}).

Equation~(\ref{resexp}) is an experimental observable that can be constructed by measuring momenta and energies alone. The integral cannot be extended  to the entire collision volume, since different fluid elements have wildly different velocities, and the analysis we have performed assumes the local rest frame of the fluid. Since the measurement is performed in the laboratory frame, we will lift this restriction in Sec.~\ref{secrelat}.

For the time being let us take a small element of fluid $\int_{\Delta V} d^3 x$ characterized by a small rapidity and transverse velocity so that the non-relativistic analysis is a reasonable starting point. 
Then integrating Eq.~(\ref{stresstovisco}) over this volume and the time duration of the particle emission $\Delta \mathcal{T}$, we have
\ba \label{obsNR}
18T \ \zeta \ \Delta V / \Delta {\mathcal T}  = \\ \nonumber
\langle \sum_{{\rm all\ }ij} \frac{(p_{ri})^2}{E_i} \frac{(p_{rj})^2}{E_j} \rangle 
- \langle N \rangle^2 \langle \frac{p_r^2}{E} \rangle^2 \\ \nonumber
+
\langle \sum_{{\rm all\ }ij} \frac{(p_{zi})^2}{E_i} \frac{(p_{zj})^2}{E_j} \rangle 
- \langle N \rangle^2 \langle \frac{p_z^2}{E} \rangle^2
\\ \nonumber
+2 \langle \sum_{{\rm all\ }ij} \frac{(p_{ri})^2}{E_i} \frac{(p_{zj})^2}{E_j} \rangle 
- 2 \langle N \rangle^2 \langle \frac{p_r^2}{E} \rangle \langle \frac{p_z^2}{E} \rangle \ .
\ea
Under the assumption of purely radial transverse flow (no vorticity) that we have invoked one can identify $p_r=p_\perp$, the perpendicular particle momentum. 

The right hand side of equation~(\ref{obsNR}) is an observable constructed with the momenta of all particles in the event database, the left hand side features the volume viscosity, that can be thus estimated if the emission time and volume (a hydrodynamic problem) and the temperature (that can be obtained by other observations such as photon or particle spectra) are known.
A theoretical observation of interest is that the event average corresponds to a microcanonical ensemble average, as the energy in the heavy ion collision is fixed while the temperature can fluctuate. However as in the next section we will keep a small volume element, we will consider its temperature as well defined, with the rest of the fluid acting as the heat bath.

One should note that the obtained fluctuations are always a lower bound since part of the particles will not be reconstructed due to detector inefficiencies, so the actual bulk viscosity will be larger than the value so obtained. Therefore, Eq.~(\ref{obsNR}) and following should be understood as $\ge$ inequalities when estimated with experimental data.

Equation~(\ref{obsNR}) can be given an alternative form in terms of each particle's energy and mass by noting that \\
$(p_r^2+p_z^2) = E^2-m^2$ as
\ba \nonumber \label{Finalrest}
18T \ \zeta \ \Delta V /\Delta {\mathcal T}  = \\ \nonumber
\langle \sum_{{\rm all\ }ij} \frac{(E^2- m^2)_i (E^2- m^2)_j}{E_i E_j}
\rangle -  \langle N \rangle^2 \langle \frac{E^2-m^2}{E} \rangle^2
\\ 
\equiv \Delta \left( \frac{E^2-m^2}{E} \right) \ .
\ea

\section{Relativistic boost of each fluid element} \label{secrelat}
 
The analysis of correlations presented hinges on Eq.~(\ref{separatensores}) that is valid in the fluid's rest frame. However, the fluid elements in the nuclear explosion following the collision are boosted in the laboratory frame. If the four-velocity of the fluid element is known to be $U^{\mu}$, Equation~(\ref{Finalrest}) can be taken to the laboratory frame 
by introducing the time-dilation factor $\gamma$ and noting that $E_i=p_i\cd U$. The result is
\ba \nonumber \label{Finallab}
18T \ \zeta \ \gamma^2 \Delta V_{\rm lab} /\Delta {\mathcal T}_{\rm lab}  = \\ \nonumber
\langle \! \sum_{{\rm all\ }ij}\!\! \frac{((p\cd U)^2\! -\! m^2)_i ((p\cd U)^2\! -\! m^2)_j}{(p\cd U)_i (p\cd U)_j}
\rangle -  \langle \! N \! \rangle^2 \langle\! \frac{(p\cd U)^2\! -\! m^2}{p\cd U} \! \rangle^2
\\ 
\equiv \Delta \left( \frac{(p\cd U)^2-m^2}{p\cd U} \right) \ .
\ea

To turn this formula into an experimentally useful expression we need to make a few more remarks. Let us assume that one has identified a set of kinematic cuts that select a swarm composed of those particles coming from the fluid element
$\Delta V_{\rm lab}$ (this will be addressed shortly) during the time interval $\Delta {\mathcal T}_{\rm lab}$.\\
The fluid element's rest frame will coincide with the center of mass frame. Therefore its velocity can be obtained from the particle swarm's energy-momentum in the laboratory frame as
\be \label{defbeta}
{\boldsymbol\beta} = \frac{\sum_i{\bf p}_i}{\sum_i E_i} \ .
\ee
Then $\gamma \equiv (\sqrt{1-\beta^2})^{-1}$ and
\be
U = \gamma (1,{\boldsymbol\beta})\ .
\ee
Once $U$ corresponding to the fluid element has been so constructed, one can compute all the products $p_i \cd U$ in Eq.~(\ref{Finallab}) as
\be
p_i \cd U = \gamma (E_i - {\bf p}_i\cd{\boldsymbol\beta}) \ .
\ee

The four-velocity $U^\mu$ satisfies $U \cdot U=1$, can be parametrized as
\be
U^\mu = \left(\sqrt{1+u_\perp^2} \cosh \eta, u_\perp \cos \phi, u_\perp \sin \phi, \sqrt{1+u_\perp^2} \sinh \eta\right) \ .
\ee 

Now let us address the fluid element's (average) space and time sizes $\Delta V_{\rm lab}$, $\Delta {\mathcal T}_{\rm lab}$ in terms of observable quantities. This requires understanding of the hydrodynamics of the expanding fireball, and here we will contempt ourselves with the simplest of models, a spherical expansion characterized by a freeze-out surface at time $\tau_f$ (this is a valid approximation if the formation radius is much smaller, $\tau_0\ll \tau_f$, else the polar caps of the sphere are distorted, and if the elliptic flow is moderately small). The total swarm's longitudinal momentum will be $p_z$ in the direction of the heavy ion beam
and is usually traded for the rapidity and the fluid element's total energy $P_z = E\ \tanh \eta$, or
$$
\eta = \frac{1}{2} \log \left( \frac{E+P_z}{E-P_z}
\right)\ .
$$
For our argument in this section 
we will consider pure radial flow, so that the swarm's perpendicular momentum ${\bf P}_\perp$ in the transverse plane is parallel to $\hat{\boldsymbol\rho}$, the radial vector in cylindrical coordinates, so the radial direction is automatically determined by the measurement of $P_\perp$ for the swarm.

We would like to express $\Delta V_{\rm lab}$ and $\Delta {\mathcal T}_{\rm lab}$ in terms of the momentum spread of the chosen particle swarm, centered around energy $E$,
 transverse momentum $P_\perp$, azimuth $\phi$ and rapidity $\eta$.

Let us start by $\Delta {\mathcal T}_{\rm lab}$. We note that in the time of freeze-out the particle travelled a distance $\rho = \tau_f \beta_{\perp}$ from the origin ($\beta_\perp = P_{\perp}/E$). A particle arriving at the freeze-out distance a time $\Delta \mathcal{T}_{\rm lab}$ later will have lagged by $\rho \Delta \beta_{\perp}$. Therefore we have
\be \label{finaltime}
\Delta {\mathcal T}_{\rm lab} = \tau_f \frac{\Delta \beta_{\perp}}{\beta_{\perp}} = \rho \frac{\Delta \beta_{\perp}}{\beta_{\perp}^2}
\ee
and differentiating $E=m\gamma=m/\sqrt{1-\beta^2}$
\be
\Delta {\mathcal T}_{\rm lab} =\frac{\tau_f}{\beta_{\perp}^2} \frac{\Delta E}{E} \frac{m^2}{E^2} \ .
\ee

Turning now to the spatial cylindrical coordinates,
\be
\Delta V_{\rm lab}\equiv \Delta z  \rho \Delta \rho \Delta \phi\ .
\ee
The longitudinal velocity gives $\Delta z = \tau_f \Delta \beta_z$.
Likewise, $\Delta \rho = \tau_f \Delta \beta_\perp$.
Altogether, employing again the definition of $\beta$ in terms of the total energy and momentum in Eq.~(\ref{defbeta}),
\be \label{finalvolume}
\Delta V_{\rm lab} = \tau_f^3 \Delta \phi \frac{P_{\perp}}{E} \left( \frac{\Delta P_z}{E}-\frac{P_z}{E^2} \Delta E \right) 
\left( \frac{\Delta P_\perp}{E}-\frac{P_\perp}{E^2} \Delta E \right)  \ .
\ee
Finally, eliminating $ \Delta P_z$ in terms of $\Delta P_{\perp} and \Delta E$, we find

\be \label{volandtime}
\frac{\Delta V_{\rm lab}}{\Delta {\mathcal T}_{\rm lab}} =
\tau_f^2 \frac{\Delta \phi}{\Delta E} \frac{P^3_{\perp}}{m^2} \left[ \left( \frac{1}{\tanh ^2 \eta}- \frac{P_z}{E}\right) \frac{\Delta E}{E} \right. \ee
\be \nonumber \left. - \frac{1}{\tanh \eta} \frac{P_{\perp} }{E^2} \Delta P_{\perp} \right]
\left( \frac{\Delta P_\perp}{E}-\frac{P_\perp}{E^2} \Delta E \right) \ . 
\ee

Note that differentiating the invariant mass of the swarm
$$
M^2=E^2-{\bf P}^2
$$
the three cuts $\Delta E$, $\Delta P_\perp$ and $\Delta \eta$ are not independent, satisfying the constraint
\be
E\Delta E = \tanh \eta E^2\Delta \eta + \cosh ^2 \eta P_\perp
\Delta P_\perp\ .
\ee

\section{Kinematic cuts}

In this section we discuss the options for the kinematic cuts, particularly 
$\Delta P_\perp$, $\Delta \phi$ that are workable for an experimental collaboration, considering especially the ALICE  experiment at the LHC.

In devising them, we have to compromise between several constraints.
\begin{itemize}
\item First, since our method calls for the separation of an interval $\Delta \mathcal{T}$
smaller than the lifetime of the entire collision, to isolate the fluctuations, we need to consider a fluid element that is actually in motion and provides us with a clock as 
in section~\ref{secrelat}. Therefore we will need to impose a $P_\perp$ cut that excludes $P_\perp=0$.
\item
Second, not all particles in a swarm move parallel enough to the average velocity $U$ and may end up in a different element of phase space.
To quantify the theory error introduced by this effect we have written a small Monte Carlo programme described shortly.
\item
Third, the phase space element chosen for the measurement needs to contain enough particles across the collision data base to make a measurement possible.
\item
Fourth and last, we have to consider that ALICE's rapidity acceptance is limited to the interval $(-1,1)$ (the barrel spans about 46 degrees in polar angle to each side of the collision point).
\end{itemize}

The crux of the matter is in the second point. The pion emission due to the freeze out of a fluid element at rest can approximately be described by a Bose-Einstein distribution in momentum $p$,
\be \label{MB1}
\frac{dN}{N} = C \frac{p^2dp}{e^{(\sqrt{p^2+m^2}-\mu)/(k_BT)}-1}
\ee
characterized by a temperature $T$ and chemical potential $\mu$.
This emission is isotropic in the rest-frame of the fluid, but if the fluid element is boosted, the boost velocity has to be compounded with the particle velocity (according to the special-relativistic velocity transformation rule). If the boost velocity is large enough, it dominates the composition. Most particles are emitted aligned with $\beta$.\\
However if the boost velocity is of order of the Bose-Einstein velocity allowed by this distribution, the emission becomes less beamed and each element of phase space is populated by particles emitted from different fluid elements.

In view of our fourth point above, since the longitudinal boost accepted by the ALICE detector has at most $\ar \eta \ar\simeq 0.9$, we will consider the central part of the collision, that is, take the entire longitudinal acceptance as one bin with $\eta=0$, $\Delta \eta\simeq 1.8$. Neglect of longitudinal momentum allows to write Eq.~(\ref{MB1}) in terms of the transverse momentum alone as
\be \label{MB2}
\frac{dN}{N} = C \frac{P_\perp^2dP_\perp}{e^{(\sqrt{P_\perp^2+m^2}-\mu)/(k_BT)}-1} \ .
\ee

To assess the kinematic cuts we proceed by writing a Monte Carlo programme. Employing Von Neumann's rejection method we generate a sample of several thousands of pions (corresponding to a few simulated collision events) distributed at random in $\phi$ and according to the ALICE  experimental $P_\perp$ distribution~\cite{appelshaeuser} in 900 GeV $pp$ collisions, that is well fit by and ad-hoc formula
\be 
\label{Alicefit}
 \frac{1}{N_{evt}} \frac{1}{2\pi P_{\perp}} \frac{d^2 N_{ch}}{d \eta dP_{\perp}} = \left\{
\begin{array}{cc} 
 11.47 \ e^{-4.10 P_{\perp}} & P_{\perp} < 1.7 \textrm{ GeV} \\
 0.25 \ P_{\perp}^{-5.95} & P_{\perp} > 1.7 \textrm{ GeV}
\end{array} \ .
\right.
\ee
This we call {\emph{defining sample}} and is only used to construct average boost velocities.\footnote{
Incidently, the same data~\cite{appelshaeuser} taken at low $P_{\perp}$ can be used to fit the rest-frame Bose-Einstein thermal distribution parameters
(temperature and pion chemical potential) in Eq.~(\ref{MB2}).}

To explore pairs of $(\Delta P_{\perp},\Delta \phi)$ cuts  
we select the pions from the {\emph{defining sample}} whose momenta fall within the so chosen fluid cell. We sum their momenta and energy to construct the cell's velocity according to Eq.~(\ref{defbeta}).

Once the fluid cell has been defined and the average velocity is known, we turn to Eq.~(\ref{MB2}) and generate a second sample of thermally distributed pions in the rest frame, also by Von Neumann's rejection method, the {\emph{thermal sample}}. \\
This sample represents isotropic emission in the fluid's rest frame and we impose no restriction on $P_{\perp}$ or $\phi$ except thermal distribution. 
\\
Finally, we apply the Lorentz boost with the velocity from Eq.~(\ref{defbeta}) corresponding to the fluid cell to each of the pions in the {\emph{thermal sample}}, and examine what fraction of them falls outside of the initial kinematic cuts that defined the fluid cell. 

We find that a non-negligible but controllable percentage of the sample pions end up into a different fluid cell. The results are listed in Table \ref{MC_results} as percentages of particles appearing with momenta that would correspond to a fluid cell other than used to generate them.

For completeness we also address ALICE's Pb+Pb data at $\sqrt{s}=2.76$ TeV.
We fit the $P_\perp$ distrbution in analogy with Eq.~(\ref{Alicefit}) by
\be \label{Alicefit2}
 \frac{1}{N_{evt}} \frac{1}{2\pi P_{\perp}} \frac{d^2 N_{ch}}{d \eta dP_{\perp}} = \left\{
\begin{array}{cc} 
  8.90 \cdot 10^5 \ e^{-2.93 P_{\perp}} & P_{\perp} < 2.0 \textrm{ GeV} \\
  2.00 \cdot 10^5 \ P_{\perp}^{-6.29} & P_{\perp} > 2.0 \textrm{ GeV}
\end{array}
\right.
\ee
to obtain the corresponding {\emph{defining sample}} and repeat the analysis (obtain each cell's velocity, generate a {\emph{thermal sample}}, boost the pions thereof and examine their final momenta). The corresponding result is given in table~\ref{MC_results2}.

\begin{centering}
\begin{table*}
\caption{We show average velocity $\beta$ of the given swarms of particles within the azimuthal angular $\Delta \phi$ and the transverse momentum $\Delta P_\perp$ kinematic cuts, with the particles distributed according to Eq.~(\ref{Alicefit}) corresponding to proton-proton collisions.
We also show, for each given binning with velocity $\beta$, the percentage of thermally emitted particles following Eq.~(\ref{MB2}) that are lost from the bin after compounding $\beta$ with the particle's thermal velocity. Typical results show that a fourth to a third of particles with well-chosen cuts populate other fluid elements introducing an irreducible theory error. \label{MC_results} }
\begin{tabular}{|c|cc|cc|cc|cc|cc|} \hline 
 & \multicolumn{2}{|c|}{All $P_{\perp}$} & \multicolumn{2}{|c|}{$P_{\perp} > 0.3$ GeV} & \multicolumn{2}{|c|}{$P_\perp > 0.5$ GeV} &
\multicolumn{2}{|c|}{ $P_{\perp} \in (0.3,2)$ GeV} & \multicolumn{2}{|c|}{$P_{\perp} \in (0.3,3)$ GeV} \\ 
\hline \hline
& $\beta$ & \% & $\beta$ & \%  & $\beta$ & \% & $\beta$ & \% & $\beta$ & \% \\ 
\hline \hline
$\Delta \phi=\pm 20 ^{\circ}$ & 0.93& 41.6 & 0.96 & 36.5 & 0.97  & 36.7 & 0.96 & 62.0 & 0.96 & 48.8 \\ \hline
$\Delta \phi=\pm 30 ^{\circ}$ & 0.91& 31.9 & 0.93 & 33.3 & 0.94 & 36.9 & 0.93 & 51.0 & 0.93 & 39.9 \\ \hline
$\Delta \phi=\pm 45 ^{\circ}$ & 0.86& 24.7 & 0.88 & 33.7 & 0.89 & 42.4 & 0.88 & 42.2& 0.88 & 35.6 \\ \hline
$\Delta \phi=\pm 60 ^{\circ}$ & 0.79& 18.6 & 0.81 & 35.7 & 0.82 & 49.2 & 0.81 & 39.6 & 0.81 & 36.3 \\ \hline 
\end{tabular}
\end{table*}
 \end{centering}

Examination of table~\ref{MC_results} teaches several general 
lessons.
\begin{itemize}
\item
If the boost velocity is generally larger (the average momentum is at higher $P_{\perp}$), pions do not spread out too much and losses from the cell are lowered. 
\item If the azimuthal-angle cut $\Delta \phi$ is larger, losses from the cell are in general smaller because, after boosting the
{\emph{thermal sample}}, most pions remain inside this larger cone.\\ 
\item If  on the other hand the azimuthal-angle cut is very small, low momentum particles find it easy to leave the  tiny resulting angular cone.
One can reduce the mixing between fluid cells by proceeding to larger $P_\perp$ so the boost focuses the swarm in the correct direction. 
 \item In the extreme case, if the momentum cut is centered at huge momenta, the cell's $\beta$ is very close to $1$.
Almost independently of the initial thermal configuration most of the particles follow the boost and fall within the {\emph{defining}} momentum cut. By increasing the angular acceptance this proportion is further improved.
However the statistics with real data fall exponentially with $P_\perp$, 
so a balance has to be found between larger momentum and sufficient data. 
(At too large momentum one should not trust thermalization either). 
\end{itemize}

  A reasonable choice would be for instance to take a small angular cut of $60 ^{\circ}$ and identify all pions with
$P_{\perp} > 0.3$ GeV. The number of particles that mix with other fluid cells is then around a third.
This mixing should be considered a systematic theory uncertainty in the measurement of the bulk viscosity.

\begin{table*}
\caption{ Same as in table~\ref{MC_results} but for lead-lead collisions, with the pion distribution following Eq.~(\ref{Alicefit2}).
\label{MC_results2}}
\begin{tabular}{|c|cc|cc|cc|cc|cc|} \hline 
 & \multicolumn{2}{|c|}{All $P_{\perp}$} & \multicolumn{2}{|c|}{$P_{\perp} > 0.3$ GeV} & \multicolumn{2}{|c|}{$P_\perp > 0.5$ GeV} &
\multicolumn{2}{|c|}{ $P_{\perp} \in (0.3,2)$ GeV} & \multicolumn{2}{|c|}{$P_{\perp} \in (0.3,3)$ GeV} \\ 
\hline \hline
& $\beta$ & \% & $\beta$ & \%  & $\beta$ & \% & $\beta$ & \% & $\beta$ & \% \\ 
\hline \hline
$\Delta \phi=\pm 20 ^{\circ}$ & 0.95& 38.1 & 0.96 & 36.1 & 0.97  & 36.6 & 0.96 & 69.9 & 0.96 & 57.2 \\ \hline
$\Delta \phi=\pm 30 ^{\circ}$ & 0.92& 30.0 & 0.94 & 33.1 & 0.94 & 36.4 & 0.94 & 59.7 & 0.94 & 47.3 \\ \hline
$\Delta \phi=\pm 45 ^{\circ}$ & 0.87& 23.9 & 0.88 & 33.2 & 0.89 & 40.9 & 0.88 & 49.9& 0.88 & 39.7 \\ \hline
$\Delta \phi=\pm 60 ^{\circ}$ & 0.80& 18.2 & 0.81 & 34.3 & 0.81 & 45.8 & 0.81 & 44.1 & 0.81 & 36.8 \\ \hline 
\end{tabular}
\end{table*}

\section{Conclusion}

We believe we have identified a way that allows to access bulk viscosity from energy-momentum correlations. 
The {\emph{modus operandi}} that we suggest would be to define three appropriate kinematic cuts $\Delta \phi$, $\Delta P_\perp$ and $\Delta E$ 
defining a swarm of particles centered around $\phi$, $P_\perp$ and $E$
to a set of recorded central collision events (many such swarms can be defined and the results compared).

Substituting then equation~(\ref{volandtime}) in (\ref{Finallab}) and evaluating the right-hand side correlator over the data, one obtains an estimate for the bulk viscosity 
\ba \label{finalbulk}  \nonumber
\zeta = 
 \frac{E^4 \Delta E m^2}{18T\gamma^2 \tau_f^2 \Delta \phi P^3_\perp} 
 \Delta \left( \frac{(p\cd U)^2-m^2}{p\cd U} \right) \times
\\  \nonumber
\frac{1}{\left[ \left( \frac{E}{\tanh ^2 \eta} - P_z \right)
\Delta E- \frac{P_{\perp}}{\tanh \eta} \Delta P_{\perp} \right] \left( E \Delta P_\perp-P_\perp \Delta E \right)} \\
\ea
that depends on the temperature $T$ and the freeze-out time $\tau_f$. These can be obtained from other measurements
 and then grant access to the bulk viscosity.  By varying the size of the cuts $\Delta E$, $\Delta \phi$, $\Delta P_\perp$ one can explore the attending systematic uncertainties, and by varying the central value of these variables around which the particle swarm is chosen, one can study the variations of the bulk viscosity over the collision volume.

Alternatively, for midrapidity ($\eta \simeq 0$) one can use the approximate formula:
\ba
\zeta = 
\frac{1}{18T\gamma^2 \tau_f^2} \frac{1}{\Delta \phi \Delta \eta} \frac{m^2}{P^2_{\perp}} \frac{E^2}{E^2- P^2_{\perp}} \Delta \left( \frac{(p\cd U)^2-m^2}{p\cd U} \right), \ \ \ \
\ea
where for the ALICE detector the longitudinal acceptance is $\Delta \eta \simeq 1.8$.
Importantly, this factor is independent of $\Delta P_{\perp}$.

One may wonder whether the bulk viscosity measured from Eq.~(\ref{finalbulk}) carries information about the entire collisions process or only the late stage
near kinetic freeze-out. Since we have chosen $\Delta \mathcal{T}_{\rm lab}$ to pick-up fluctuations in Eq.~(\ref{finaltime}), the measurement refers to precisely
the interval $\Delta \mathcal{T}_{\rm lab}$ before freeze-out.
For exmaple if $\Delta E \sim 2$ GeV, $E \sim 500$ MeV, $m \sim 140$ MeV and $\tau_f \sim 10$ fm/$c$, then $\Delta \mathcal{T}_{\rm lab}$ refers to the
last $3$ fm/$c$ before the freeze-out.


\vspace{2cm}
{\emph{ 
We thank P. Ladr\'on de Guevara for his feedback on the experimental feasibility of the measurement by the ALICE collaboration.
Work supported by grants FPA 2008-00592, FIS2008-01323 plus 227431, Hadron-
Physics2 (EU) and PR34-1856-BSCH, UCM-BSCH GR58/08, 910309, PR34/07-
15875. JMT is a recipient of an FPU scholarship.
}}

\appendix


\end{document}